# ROAD ACCIDENT PREVENTION UNIT (R.A.P.U)
## (A Prototyping Approach to Mitigate an Omnipresent Threat)

By Dibakar Barua, Pranshu Jain, Jitesh Gupta, Dhananjay V. Gadre
Dept. of Electronics and Communication

Netaji Subhas Institute of Technology, New Delhi -78
dibakar.barua92@gmail.com

**Abstract**

Road accidents claim a staggeringly high number of lives every year. From drunk driving, rash driving and driver distraction to visual impairment, over speeding and over-crowding of vehicles, the majority of road accidents occur because of some fault or the other of the driver/occupants of the vehicle. According to the report on "Road Accidents in India, 2011" by the Ministry of Transport and Highways, Government of India, approximately every 11th person out of 100,000 died in a road accident and further, every 37th person was injured in one, making it an alarming situation for a completely unnecessary cause of death.

The above survey also concluded that in 77.5 percent of the cases, the driver of the vehicle was at fault. The situation makes it a necessity to target the root cause of road accidents in order to avoid them. While car manufacturers include a system for avoiding damages to the driver and the vehicle, no real steps have been taken to actually avoid accidents. "Road Accident Prevention Unit" is a step forward in this stead. This design monitors the driver's state using multiple sensors and looks for triggers that can cause accidents, such as alcohol in the driver's breath and driver fatigue or distraction. When an alert situation is detected, the system informs the driver and tries to alert him. If the driver does not respond within a stipulated time, the system turns on a distress signal outside the vehicle to inform nearby drivers and sends a text message to the driver's next of kin about the situation. A marketable design would also shut down power to the vehicle, thus providing maximum probability for avoiding road accidents and extending a crucial window for preventive and mitigation measures to be taken.

**1. Introduction**

According to the study conducted by the Ministry of Transport and Highways, only 9 percent of the accidents observed were attributed to material causes such as faults in the road, weather conditions, vehicular defects etc. and a meager 3.7 percent of the accidents were caused when a cyclist or pedestrian was at fault. Further, 60 percent of the driver-caused road accidents were attributed to over speeding, 16.7 percent of these were due to alcohol or drug consumption and lastly 23.6 percent were caused due to driver fatigue or overcrowding of vehicles. These clearly bring to light the gravity of the situation and the enormous responsibility of vehicle drivers towards causing road accidents. "R.A.P.U" is motivated by the desire to curb such incidents which are caused due to a moment of madness or complete irresponsibility of the driver as such situations are easily avoidable. A life lost in a road accident is unforeseen and absolutely unnecessary, making the addressing of the situation a complete must. The technical content of R.A.P.U's design is inspired by the module suggested in "Accident Prevention Using Eye Blinking and Head Movement", published in the International Journal for Computer Applications® (IJCA) [1]. Our goal was to make the suggested design simpler to implement, low cost, low power, easily installable and extendable. R.A.P.U has been dealt with in an entry level manner and has been tested for premium effectiveness.

**1.1 Technical Background**

This issue has previously been dealt with in quite detail in "Accident Prevention Using Eye Blinking and Head Movement", published in the International Journal for Computer Applications® (IJCA) [1].

The author suggested the use of spontaneous video capture and image processing to analyze the driver's state. However the key factors inhibiting the driver's ability i.e. eye blinking and head tilting can be easily analyzed using low cost sensors such as accelerometers and IR sensors, thus obviating the need for expensive equipment as well as advanced processing platforms. Including these in the "Road Accident Prevention Unit" has made the design a low cost, low power implementation which can be easily installed and changed. The accident prevention and mitigation steps involve a Distress Signal visualization outside the vehicle using a Display Board, alarm intimation for the driver and a text message sent to the next of kin with the GPS coordinates of the driver upon non-attendance of the situation.

**1.2 What does R.A.P.U do?**

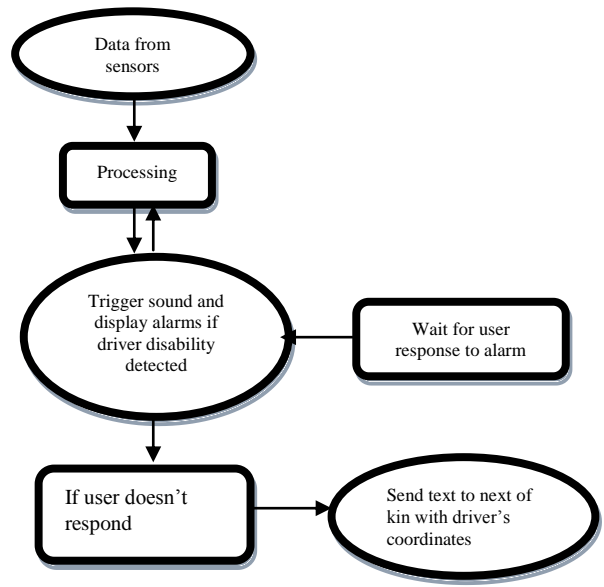

**Figure 1:** R.A.P.U, a bird's eye view

Figure 1 shows a top level block diagram of the entire system. The key responsibility of the design is to constantly fetch data from the sensors regarding the driver's ability to drive safely. If the system renders the driver incapable to do so, appropriate steps are taken to notify the driver's nearby vehicles and the driver's next of kin of the situation so that suitable prevention and mitigation steps can be taken. The design has been made in a modular approach in such a way that it can be easily installed in any vehicle. The modules can be individually installed as per the need of the user. Overall the system is a low cost and low power solution to the issue which has enormous potential for adaptation and expansion to be made road-worthy.

**1.3 Organization of the report**

The paper describes the technical aspects of R.A.P.U in detail. The top level hardware and software modules are described in Section 2 along with the approach taken to implement the system.

Section 3 deals with the Hardware and Software design of the system extensively. The various analog components in the front as well as back end are explained. The software algorithms and interfacing with the various modules of the system are also explained.

Section 4 details the conclusions and results of the tests that were carried out on R.A.P.U. Tests were carried out on multiple subjects in different environments and the results of the trials have been suitably drafted. Software trials of the design have also been explained.

The future scope of R.A.P.U, which is essentially its strong point, and its limitations and issues have also been dealt with.

The report ends with the details of the PCB designs and the software model that have been used in the system, provided in Appendix A and B.

**2. Proposed Solution**

**2.1 Hardware Components**

The hardware model of R.A.P.U has been broadly divided into the following modules:

a. Sensor Interface proto-shield for attaching all peripherals to microcontroller unit.
b. Sensor and Alarm Platform which has all the sensors and a speaker unit to be mounted on the driver.
c. Display Unit to be mounted outside the vehicle.
d. Amplification and filtering Unit for sending enlarged Alarm Sound to the speaker's Sensor framework.
e. LCD unit for visual interface to the system.
f. Communication Unit for using the GSM and GPS modules to send a text message.
g. Power Supplies for powering the entire device and the GSM module.

A few assumptions have been made with regard to the hardware implementation of R.A.P.U:

a. The IR sensor has been calibrated for the ambient setting before using the device.
b. The Sensor unit does not cause any interference with the driver's ability.
c. The Display Shield is visible to the drivers nearby.
d. The slight delay between Alarm Trigger and message being received by the next of kin is inconsequential and of no significant value.

e. The power supplies designed can be easily modified to draw power from the vehicle.

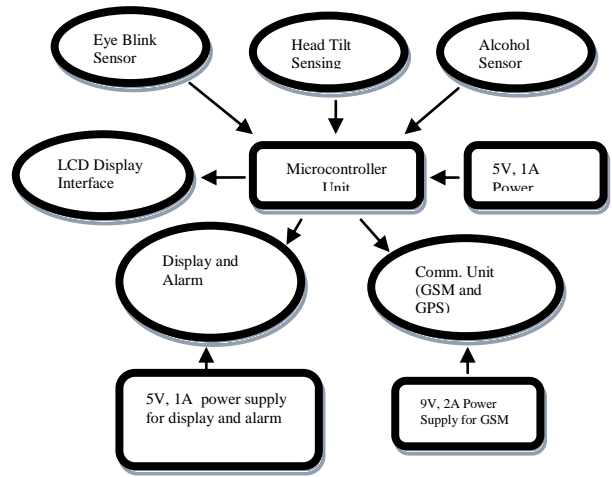

**Figure 2:** Hardware Model

There are certain constraints that R.A.P.U's hardware design inflicts upon the user. Firstly the user has to wear the sensor unit in order to use the system which is an added responsibility along with the vehicle. The Communication Unit has a fixed contact number programmed into it to which the message will be sent. Also, the GSM unit uses a larger power supply separate from the system.

The analog components of the system are involved in the backend of the hardware design. The Infra Red sensor utilizes an LM358 op-amp for amplification of the signal so that it can be calibrated to adjust to ambient light settings. The Speaker unit uses an Amplification Unit which uses an audio grade LM386 op amp for amplification and filtering. The primary power supply uses an LM723 Voltage Regulator by Texas Instruments to provide a stable 5V, 1A power supply.

**2.2 Software Components**

The Software model of the design consists of the following:

1. Display of system status on the Device Liquid Crystal Display.
2. Auto Calibration of Accelerometer to adjust to driver position upon System Reset.
3. Reading the sensor data in a polled fashion.
4. Triggering the Alcohol Alarm if alcohol is detected in driver's breath.
5. Triggering the Fatigue Alarm if Head Tilt or Eye Blinks are detected.
6. Triggering the Communication Unit to send a text message with GPS coordinates as final steps.

**3. Implementation**

**3.1 Hardware Implementation**

The various Hardware modules used were:

a. Sensor Unit: comprising of a MQ2 alcohol and combustible gas sensor, an MMA7760 Low-G accelerometer and an Infra Red Sensor Breakout Board consisting of an LM358 Op Amp by Texas Instruments and a potentiometer for calibration.

b. Alarm Device consisting of a Display Board made using Seven Segment Displays and a 5V relay and a Speaker Unit consisting of two 4 Ohm Speakers, a amplification and filtering unit based on the audio grade LM386 Op Amp by Texas Instruments and a 3.5 mm jack based Headphone unit.

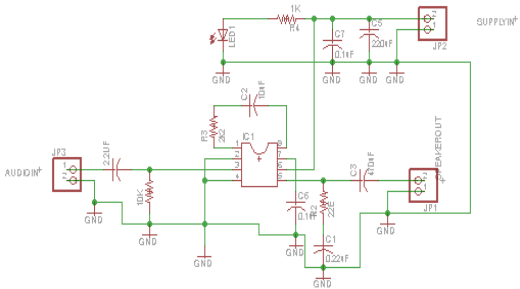

**Figure 3:** Schematic for Sound Amplifier

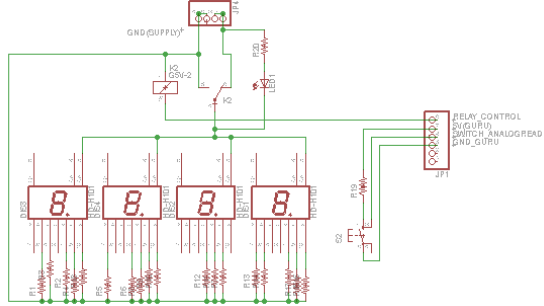

**Figure 4:** Schematic for Display

c. LCD Unit consisting of an HD44780 based 16X2 Liquid Crystal Display as a System Interface.

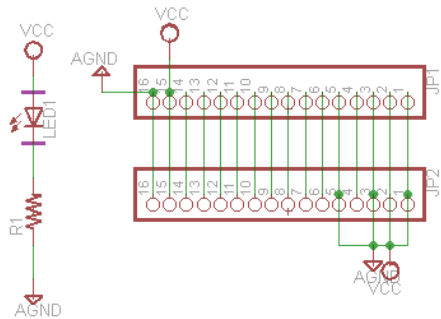

**Figure 5:** Schematic for LCD Board

d. Communication Unit consisting of a GNS TC6000-GN GPS Unit by Texas Instruments, a SIMCOM GSM Module along with a 9V, 2A Power Supply and an AT Mega 328 based microcontroller board.

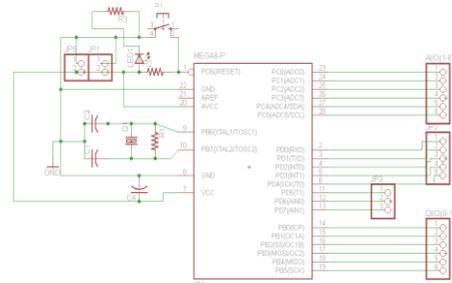

**Figure 6:** Schematic for Communication Unit

e. A 5V, 1A Power Supply based on the LM723 voltage regulator by Texas Instruments.

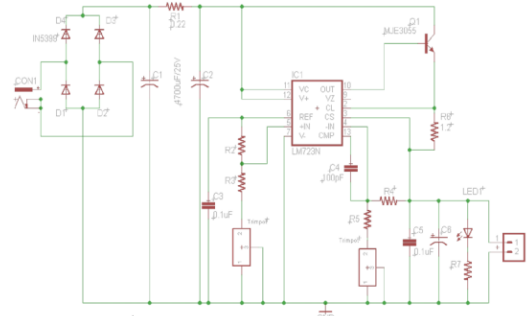

**Figure 7:** Schematic for Power Supply

f. The Microcontroller Unit comprising of the Stellaris Guru Evaluation Kit by Texas Instruments based on LM3S608 MCU and a proto-shield for peripheral interfacing.

The various integrated circuits that have been used are:

a. LM358 Op Amp: has been used in the IR Sensor Board for amplification of the sensed voltage for adjustment using a potentiometer. This serves as a calibration for adjusting to the ambient lighting. This IC was particularly chosen due to its single 5V supply requirement which was easily provided by our microcontroller.

b. LM386 Op Amp: has been used in the Audio Amplifier circuitry. This IC has been chosen due to its audio grade properties and single supply requirement which helps provide a distinguished sound output.

c. LM723 Voltage Regulator: has been used in the 5V, 1A Primary Power Supply. This has been used due to its ready customizability by changing the output voltage by simply changing the value of a potentiometer.

d. LM3S608 Stellaris Cortex M3 Microcontroller: has been used in the Stellaris Guru Evaluation Kit as the primary processor. The Stellaris Family has been chosen due its fast processing capability, Low Power working and sufficient memory provisions which adhere appropriately to the system design.

e. AT Mega 328 Microcontroller: has been used in the Communication Module for easy interfacing of the GSM and GPS kits which require a 5V-GND digital communication protocol.

**3.2 Software Implementation**

The software model follows a modular approach consisting of the following components:

a. The LCD Display Library written exclusively for the Stellaris Cortex-M3 Peripheral Driver Library
b. The Auto Calibration of the Accelerometer Reference upon System Reset.
c. Sensor reads in polling fashion.
d. The Alarm State for fatigued state which can be escape within a period of 10 seconds using a button which is serviced as a hardware interrupt.
e. The Speaker and Relay Control using the Stellaris Driver Library achieved by configuration of the GPIO output current set to 8mA.

The software model is best described by the following framework diagram:

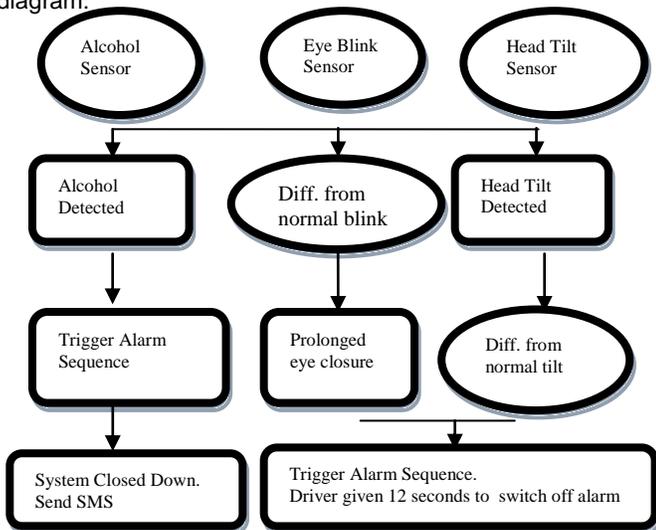

**Figure 8:** Software Framework Overview

As shown, the sensors are read in a polled fashion with the alcohol sensor readings checked first. Even one trigger in this case locks the system down and sends the SMS. The eye blinks and head tilts are differentiated from arbitrary head movements and eye blinks by re-reading of the sensors upon detection.

The device uses an Auto Calibration Algorithm upon System reset to learn the the driver's head's reference coordinates as per his seating configuration. The deviations from these values are considered for detecting his head movement.

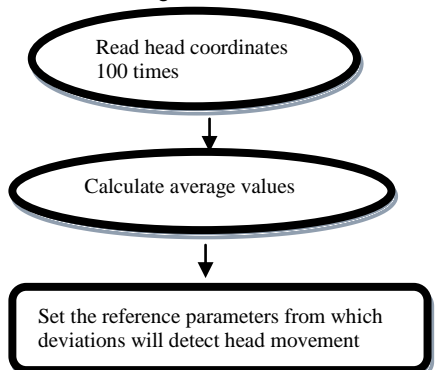

**Figure 9:** Auto Calibration Procedure for Driver's Head Configuration

The Alarm and Display Board of the system use a 5V relay which is controlled by the 3.3V GPIO output of the microcontroller using a software-modified pad current of 8mA. The Push Button which allows the driver to escape the Alarm Sequence is configured as a Hardware Interrupt Controlled GPIO Input for immediate redressal and to ensure the driver only has to press it once at any time once the Alarm is triggered.

**4. Results**
The system devised is a Low Power and Easily Installable Modular device. The hardware components were tested for maximum Power Consumption. R.A.P.U has an absolute maximum current rating of 38mA when all the GPIO Outputs are loading the processor working at a voltage of 3.3V providing an ultra low power treatment.
The device achieved a user-independent compatibility for eye blink and head tilt detection through rigorous testing to determine the most appropriate comparison counters.

1. **Eye Blink Detection**: The I.R Sensor used for Eye Blink Detection uses an I.R transmitter for throwing I.R light at the user's eye. The open eye absorbs this light while the closed eye reflects it, thus allowing us to determine the eye state. This mechanism is adjustable using a potentiometer to ensure that the ambient lighting does not affect the device working. This was proved to be a better alternative than Image Processing as camera readings are light dependent and the processing requirements are also less in R.A.P.U.
The device uses a re-reading mechanism to ensure that stray blinks to not trigger the system. It was determined that a closed reading of 12 values out of 15 resembled an eye that was closed for approximately 2-3 seconds in our detection scheme.

2. **Head Tilt Detection:** The device measures deviations from the reference head coordinates, which are a learned upon System Reset, to determine head movement. A significant deviation triggers a re-reading mechanism to fully ascertain the head movement to ensure that stray movements such as on vehicle turns, music etc. are not detected.

**5. Conclusions**

R.A.P.U is an easy-to-install system that can be implemented in most situations. However if the driver actually fails to wear the Sensor Framework Unit, the device cannot be used. Also there is always the issue of adjustment of the calibration potentiometer in the I.R Sensor to ensure the ambient light does not interfere with the sensing process.

R.A.P.U's true potential lays in the future implementations which are possible to make the system a complete accident detection and prevention unit.
1. **Automatic braking system** (to slow down the speed of car) or lock down of car (in case of alcoholic state) can also be implemented to improve the efficiency of RAPU.
2. To monitor the **pulse rate** of the driver to observe any sudden change so that any situation of panic can also be pre-detected.
3. The sensor framework can be made a **Wireless** node using a secondary processor and sensors to achieve a complete modular system.
4. Provisions can be made for detecting **Rash Driving** using a large number of sensors and suitable sensing algorithms.

**5. R.A.P.U As Mobile Based Device**

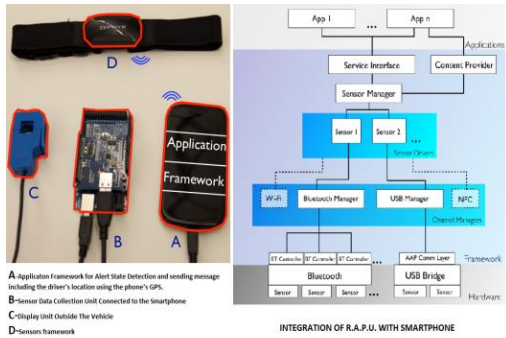

A - Application Framework for Alert State Detection and sending message including the driver's location using the phone's GPS.
B - Sensor Data Collection Unit Connected to the Smartphone
C - Display Unit Outside The Vehicle
D - Sensors framework

INTEGRATION OF R.A.P.U. WITH SMARTPHONE

R.A.P.U will be a complete and ubiquitous system if the device is configured as a Mobile Based Framework which configures itself to the user's smartphone using Bluetooth. Not only will this obviate the need for expensive GSM and GPS Units but also allow the device to use the mobile's connectivity to achieve instant reactivity to Trigger Situations. Android applications for the User End and the Traffic Controller/ Police Booth End will allow the device to upload the details of an accident-prone scenario directly to the corresponding servers to achieve immediate addressal of such a scenario.

**References**

1. Accident Prevention "Using Eye Blinking and Head Movement", by Abhi R. Varma, Chetna Bharti and Seema V. Arote, Proceedings published in International Journal of Computer Applications® (IJCA), Emerging Trends in Computer Science and Information Technology -2012(ETCSIT2012).

2. "Head Mounted Input Device using MEMS Sensors", by Anbarasu V (Research scholar, Sathyabama University) and Dr. T. Ravi (Professor & Head, Dept of CSE, KCG College of Technology), Proceedings Published in Indian Journal of Computer Science and Engineering (IJCSE)

3. "Accident Avoidance and Prevention on Highways" S.P Bhumkar, V.V Deotare, R.V Babar , Singhad Institute of Technology, Lonavala, Processidings published in International Journal of Engineering Trends and Technology (IJETT Vol. 3, Issue 2, 2012)

**APPENDIX C: Bill of Materials**

| | Component | Manufacturer | Cost per component | Quantity | Total cost of component | TI Supplied/ Purchased |
|---|---|---|---|---|---|---|
| 1. | LM 723 Voltage Regulator | TI | Free | 1 | Free | TI |
| 2. | LM3S608 MCU Unit on ARM Stellaris GURU | TI | Free | 1 | Free | TI |
| 3. | LM358 op amp | TI | Free | 1 | Free | TI |
| 4. | LM386 op amp | TI | Free | 1 | Free | TI |
| 5. | CC4000 GPS Module | TI | Free | 1 | Free | TI |
| 6. | MMA7760 Acceler-ometer | Freescale | Rs. 150 | 1 | Rs. 150 | Purchased |
| 7. | IR leds | Generic | Rs. 20 | 4 | Rs. 80 | Purchased |
| 8. | 10K pot | Generic | Rs. 10 | 5 | Rs. 50 | Purchased |
| 9. | SIMCOM GSM module | Robokits | Rs. 2000 | 1 | Rs. 2000 | Purchased |
| 10. | DC adapters | Generic | Rs. 50 | 2 | Rs. 50 | Purchased |
| 11. | Pinheads Jumpers LEDs Capacitors and other electronic components | Generic | Rs. 300 | - | Rs. 300 | Purchased |
| | | | Total cost of the project | | Rs. 2580 | |